\documentclass[twocolumn,9pt]{article}
\usepackage[margin=1in]{geometry}
\usepackage{newtxtext}
\usepackage{newtxmath}

\usepackage{amsmath,amssymb,amsfonts}
\usepackage{graphicx}
\usepackage{booktabs}
\usepackage{multirow}
\usepackage{url}
\usepackage{tabularx}
\usepackage{hyperref}
\usepackage{natbib}
\usepackage{orcidlink}

\hypersetup{
    colorlinks=true,
    linkcolor=blue,
    citecolor=blue,
    urlcolor=blue,
    breaklinks=true
}

\title{Using AI for User Representation: An Analysis of 83 Persona Prompts}

\author{
    Joni Salminen\orcidlink{0000-0003-3230-0561}\\ 
    University of Vaasa\\
    Vaasa, Finland\\
    \texttt{jonisalm@uwasa.fi}
    \and
    Danial Amin\orcidlink{0009-0000-7597-2267}\\
    University of Vaasa\\
    Vaasa, Finland\\
    \texttt{danial.amin@uwasa.fi}
    \and
    Bernard J. Jansen\orcidlink{0000-0002-6468-6609}\\
    Qatar Computing Research Institute\\
    Hamad Bin Khalifa University\\
    Doha, Qatar\\
    \texttt{jjansen@acm.org}
}

\date{}

\begin{document}

\maketitle

\begin{abstract}
We analyzed 83 persona prompts from 27 research articles that used large language models (LLMs) to generate user personas. Findings show that the prompts predominantly generate single personas. Several prompts express a desire for short or concise persona descriptions, which deviates from the tradition of creating rich, informative, and rounded persona profiles. Text is the most common format for generated persona attributes, followed by numbers. Text and numbers are often generated together, and demographic attributes are included in nearly all generated personas. Researchers use up to 12 prompts in a single study, though most research uses a small number of prompts. Comparison and testing multiple LLMs is rare. More than half of the prompts require the persona output in structured format, such as JSON, and 74\% of the prompts insert data or dynamical variables. We discuss the implications of increased use of computational personas for user representation.\footnote{This paper has been accepted at the 22nd ACS/IEEE International Conference on Computer Systems and Applications AICCSA 2025.}
\end{abstract}


\section{Introduction}

User personas (``personas'', for short) are fictitious user representations based on real data about a user population, such as users of a system system or customers of a product \cite{cooper_inmates_1999}. Personas aim to represent user segments with distinct needs, wants, and circumstances, so designers or other stakeholders can make more informed decisions about the people that personas represent \cite{nielsen_personas_2019}. This logic is part of user-centered design (UCD), which is a subfield of human-computer interaction (HCI), though personas have permeated other domains as well, including marketing and business, health informatics, sustainability, and so on \cite{salminen_use_2022}. 

Although user data, such as interviews, have traditionally been analyzed by human persona creators \cite{nielsen_personas_2019}, computational techniques also have a substantial tradition in persona creation  \cite{jansen_data-driven_2021}. Specifically, data-driven personas tend to imbue algorithms, machine learning (ML), artificial intelligence (AI), and data science to quantitatively analyze user data collected from online sources in real-time or periodically, often using application programming interfaces (APIs) \cite{jung_automatically_2018}. Systematic reviews in the field \cite{salminen_survey_2021} reveal persona development (or generation, which we use interchangeably with persona development or creation) has co-evolved together with technical progress, from user segmentation algorithms to web-based interfaces \cite{jung_automatically_2018}, and, more recently, large language models (LLMs) and other techniques of generative AI (GenAI) \cite{jung_personacraft_2025,shin_understanding_2024}. 

The technical evolution of persona development techniques promises advantages for faster, more efficient, and more accurate user representation through the use of effective and efficient computational models \cite{jansen_data-driven_2021}, such as LLMs. However, the process of applying technology in persona development has always come with a risk, ranging from algorithmic bias to problems in training data and the general statistical issues with factors like central tendency when trying to capture diverse user populations \cite{prpa_challenges_2024,baki_kocaballi_conversational_2023}. So, researchers are very aware that ``things can go wrong'' when applying sophisticated technologies in user representation, particularly persona development. 

To mitigate these potential harms, key factors for ``good'' persona development are that (a) personas should be based on primary data about the user population they aim to represent and (b) their development process should be logically justifiable and empirically scrutinizable \cite{chapman_personas_2006,salminen_deus_2024}. While such guidelines exist for the application of computational techniques in persona development in general, specific techniques, such as the use of LLMs, often warrant their own, dedicated guidelines; the ``devil is in the details'' of applying these methods in a safe and productive manner. 

To this end, the current body of knowledge is lacking multiple contributions, including (a) systematic mapping of how researchers apply LLMs in persona generation and (b) empirical studies on the effects of these choices on the outcome personas. Without these inquiries, it will be difficult, if not impossible, to provide evidence-based guidelines on the appropriate use of knowledge technologies, specifically LLMs and GenAI, in persona development. 

More specifically, LLMs operate by taking prompts; that is, instructions from people on what these people wish the LLMs to output. Because they are based on natural language, the design space for prompts is enormous---one can literally instruct an LLM to generate personas in a close-to-infinite number of ways. So, \textit{what is the correct way? What is the best way? What are the ways that definitely should not be applied, because they cause harm?}

These are some of the motivational questions inspiring our current work, which deals with a structured analysis of \textit{persona prompts} that we have extracted from literature that uses LLMs in persona development or evaluation. Persona prompt refers to this precise practice of using LLMs (through prompting) for persona development or evaluation, instructing the LLM to generate personas either partially or entirely. We believe that a systematic mapping and analysis of persona prompts currently prevailing in the literature is a worthwhile research goal, as it provides useful information for research colleagues and the community on this evolving practice of using LLMs productively and safely for persona generation.

To this end, we pose the following research questions (RQs):

\begin{itemize}
    \item \textbf{RQ1:} \textit{Why do researchers use persona prompts?}
    \item \textbf{RQ2:} \textit{How do researchers use persona prompts?}
    \item \textbf{RQ3:} \textit{What kind of personas do researcher generate with persona prompts?}
\end{itemize}

We address these RQs by analyzing persona prompts introduced and applied by researchers. 

\section{Related Work}

GenAI and LLMs have permeated the data-driven persona development primarily in three ways: (1) as one-shot exercises to generate personas with or without primary user data \cite{de_paoli_writing_2023, salminen_deus_2024}, (2) to complete a series of tasks in the overall persona creation process \cite{jung_personacraft_2025}, and (3) as a form of human-AI collaboration to support design tasks \cite{shin_understanding_2024}. For example, De Paoli demonstrated how to perform a thematic analysis to generate personas from semi-structured interviews \cite{de_paoli_writing_2023}. Jung et al. \cite{jung_personacraft_2025} integrated LLMs into an interactive persona system to address four types of tasks: (a) quantifying user attitudes, (b) identifying themes in open-ended responses, (c) grouping and summarizing survey questions, and (d) generating persona narratives including headlines and descriptions. The key argument was that LLMs perform reasonably well with tasks traditionally handled by humans, such as persona template labeling and narrative writing \cite{jung_personacraft_2025}. Shin et al. \cite{shin_understanding_2024} investigated collaborative workflows for persona generation, finding that personas generated in collaboration between LLMs and humans evoke more empathy than personas generated by either party independently.

A key question in applying LLMs in persona development relates to prompt design. While there are guidelines for prompting \textit{in general} \cite{schulhoff2024prompt,wang2024prompt}, thus far, there appear to be persona-specific instructions. While some general guidelines, such as setting a system role \cite{zheng2024helpful} or defining output requirements \cite{liu2024llms}, may apply, persona development is a specialized user representation task, implying that prompt design for persona development requires its own set of best practices. For example, Sun et al. \cite{sun_persona-l_2024} discuss (1) role-play prompting, (2) one-shot prompting, and (3) incremental prompting. 
In \textit{role-play prompting}, the LLM takes the role of the persona\footnote{For example, ``You are Shea. You are a 33-year-old woman with Down syndrome, bursting with enthusiasm and confidence.'' \cite{sun_persona-l_2024}.}. \textit{In one-shot prompting}, the LLM is given an example of a persona to ensure its understanding of the structure for the desired persona information\footnote{For example, ``Target Group: People involved in Down syndrome. Name: Shea. Age: 33. Profession: Retail associate. Hobby + interest: Passionate about disability advocacy, public speaking, and personal growth...'' \cite{sun_persona-l_2024}.}. In \textit{incremental prompting}, the persona information is gradually expanded based on internal or external inputs\footnote{For example, ``(1) Prompt: Create Personas (2) Prompt: Add the following attributes for each previously created persona: $<$keywords$>$'' \cite{sun_persona-l_2024}.}. 

In conclusion, prompts have been ingrained into current persona development practices. However, as far as we know, there is no systematic analysis of these prompts and their characteristics. We postulate that this forms a major research gap in data-driven persona research that should be addressed to provide (a) situational awareness of the current usage of prompts in personas, and (b) a platform for more formal empirical analyses on the effects of prompting on the generated personas, behavior of stakeholders using LLM-generated personas, and a fair assessment of risks and opportunities in the integration of LLMs and GenAI into persona development. To this end, we analyze 27 persona prompt entries that we extracted from research articles.

\section{Methodology} \label{method}

The prompts analyzed in this study are systematically extracted from a corpus of 52 articles identified through the systematic literature review (SLR) that examined GenAI applications in persona development \cite{amin_how_2025}. The complete methodology for the identification and selection procedures of the articles is detailed in related work \cite{amin_how_2025}, which serves as the basis for the current prompt-focused analysis. However, a summary of the same process is presented here as well. Building upon this systematically selected corpus, the present study extends the investigation by conducting a detailed examination and coding of the prompting strategies used in these research contributions.

\subsection{Article Extraction}
The articles are derived from the SLR that examined how GenAI is used in persona development \cite{amin_how_2025}. The authors in the SLR searched six academic databases, namely the ACM Digital Library, ACM Guide to Computing Literature, IEEE Xplore, Web of Science, Scopus, and arXiv to find the relevant articles. The search query used a combination of  GenAI terms (such as ``generative artificial intelligence'',``large language model'',``GPT'',``ChatGPT'', ``Claude'', and ``LLAMA'') with persona-related terms (including ``data-driven persona'',``user persona'',``persona development'',``persona creation''). The initial search identified 573 articles, which were subsequently screened using the PRISMA guidelines. After removing duplicates and applying inclusion criteria of including only peer-reviewed and pre-print articles (that are less than 2 years old) that are using Generative AI to develop and use a persona in the HCI context, the corpus was reduced to 52 articles. This also included three articles collected using the snowball technique.

\subsection{Prompt Extraction}
During the coding of the articles in the SLR, the research team systematically examined all 52 articles to identify and document prompts used in different stages of persona lifecycle. Prompts were available in almost half (n = 27, 52\%) of the articles. Extraction focused on direct prompt text, descriptions of their usage with GenAI models, and any prompting techniques they described. When an article contained multiple prompts or prompt variations, these were maintained as single entries to preserve the complete picture of researchers approach prompting. For the extraction, only the article text and appendices were considered. If an article containing multiple prompts of a system, in which persona generation is only a component, the prompts associated with persona generation were retained. The final list includes 
27 prompt \textit{entries}. As each entry can include multiple prompts (researchers typically use more than one prompt in persona generation), the total evidence base we analyze is 83 persona prompts (available the online appendix\footnote{\url{https://osf.io/rvnax/?view_only=898e41f4bd47447583517dba87a73d02}}).

\subsection{Coding}

To address our RQ, we developed a coding framework that provides the necessary information (available the online appendix\footnote{Ibid.}). Coding was performed collaboratively by two of the authors, and each prompt was analyzed according to the established framework. During the coding process, in addition to the prompt text, the manuscripts were also reviewed to extract relevant information and verify the accuracy of the prompts. To ensure reliability of the coding process, the two authors randomly cross-checked each other's work and any disagreements were discussed to decide if changes were necessary for answering the RQs. 


\section{Results} \label{results}

\subsection{RQ1: Why Do Researchers Use Persona Prompts?}

Researchers use LLMs to create, evaluate, and apply personas across diverse applications. Among others, these applications include using personas as educational tools for training counselors \cite{rudolph_ai-based_2024}, as proxies for audience understanding \cite{choi_proxona_2024}, and as instruments for storytelling and character development \cite{park_character-centric_2024}. Researchers also investigate methodological considerations for persona creation \cite{paoli_improved_2023}, evaluate the quality and limitations of LLM-generated personas \cite{schuller_generating_2024}, including potential biases and ethical concerns in these representations \cite{gupta_evaluation_2024, sun_persona-l_2024}, and investigate how visual elements affect user perception \cite{salminen_picturing_2024}. 

Specific application areas cover a broad range of topics, ranging from climate change communication \cite{nguyen_simulating_2024} to design and commercial applications \cite{goel_preparing_2023}, which reflects the diversification of persona application into societal domains beyond software development and technology \cite{salminen_use_2022}. 


The use of LLM for personas focuses primarily on persona generation (n = 22, 81.48\%), with fewer studies addressing prediction with personas (n = 7, 25.93\%) or persona evaluation (n = 4, 14.81\%). However, prediction does impose a relatively frequent use case, given that a quarter of the coded instances apply the persona concept for prediction; this took place, e.g., to predict the persona's numerical answers \cite{clements_innovative_2023} or determine compatibility between online content and a set of personas \cite{choi_proxona_2024}. Prediction marks an interesting use case for personas, because the previous generation of data-driven personas focused on generating personas without predictive capabilities \cite{jansen_flat_2020}, even though persona users were provided with limited interaction techniques, such as generation, search, and filtering \cite{jung_automatically_2018}. In turn, researchers are now using LLM-generated personas to generate predictive responses in various scenarios \cite{nguyen_simulating_2024,li_steerability_2024}. 

Evaluation approaches ranged from straightforward coherence ratings to complex multi-step evaluations that analyzed how well personas maintained consistency across different contexts and interactions. Specific evaluation techniques included (1) embedding personas in follow-up story generation tasks and implementing self-evaluation mechanisms to assess integration and consistency within fictional storylines \cite{park_character-centric_2024}, (2) analyzing coherence, content accuracy, and conversation flow in counseling contexts using a structured JSON-based rating system \cite{rudolph_ai-based_2024}; (3) measuring interactive personas using Likert scales for fluency and helpfulness while also incorporating free-form feedback by the evaluator LLM \cite{li_iqa-eval_2024}; and (4) evaluating how well personas reflected YouTube audience characteristics through classification of comments and dimension-value sets, such that the LLM first uses dimension-value pairs to classify audience comments, and then evaluates persona consistency by checking if the generated personas appropriately reflect these classified characteristics \cite{choi_proxona_2024}.




Slightly more than half (n = 14, 51.85\%) of the prompt entries required a structured output, while the rest posed no structure requirements (n = 13, 48.15\%). When a structured output is required, it is primarily in JSON\footnote{``Provide the output in a json array, with each dict containing only the following keys: `index', `name', `age', `occupation', `background', `details';'' \cite{salminen_deus_2024}.} (n = 7, 50\%), with other formats (n = 7, 50\%), such as tables. Structured output is particularly useful for downstream tasks like data analysis \cite{cheng_marked_2023}; unstructured format suits persona narratives \cite{chen_why_2025} and dialogue with persona users. In particular, several researchers deployed personas to study LLMs, e.g., from the perspective of bias \cite{gupta_evaluation_2024} or correspondence with actual user populations. It is, therefore, common that the prompts employ structured outputs (including JSON formats) to systematically rate personas.

\subsection{RQ2: How Do Researchers Use Persona Prompts?}

\begin{figure*}
    \centering
    \includegraphics[width=0.7\linewidth]{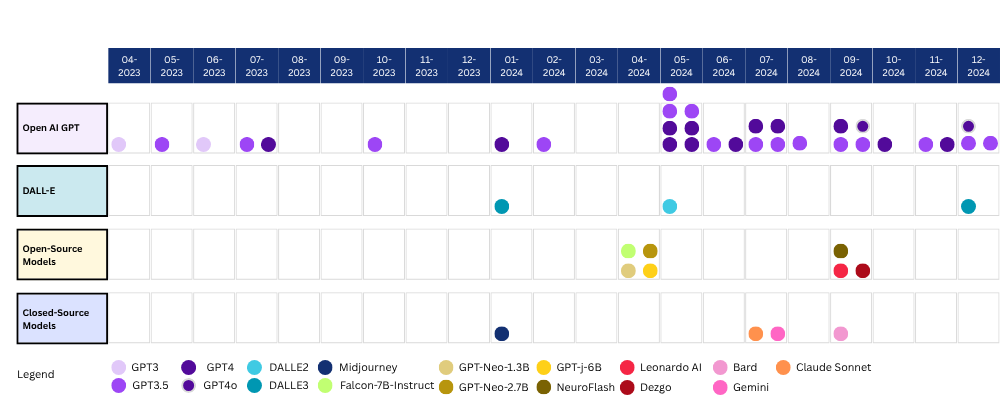}
    \caption{Evolution of the LLMs used for persona generation.  We encourage the reader to zoom in if text is hard to read.}
    \label{fig:llms}
\end{figure*}

The overwhelming majority (e.g., \cite{chen_why_2025,rudolph_ai-based_2024,clements_innovative_2023, schuller_generating_2024}) of model instances used GPT models (n = 35, 76.1\% of all model instances). The only other model family to appear multiple times was DALL-E (n = 3, 6.5\%) (e.g., used in \cite{salminen_picturing_2024,park_character-centric_2024}, to generate persona images), while every remaining model (Midjourney \cite{salminen_picturing_2024}, Bard \cite{schirgi_development_2024}, Claude \cite{gupta_evaluation_2024}, Gemini \cite{gupta_evaluation_2024}, Falcon \cite{li_steerability_2024}, GPT-Neo \cite{li_steerability_2024}, GPT-j \cite{li_steerability_2024}, NeuroFlash \cite{schirgi_development_2024}, Leonardo AI \cite{schirgi_development_2024}, and Dezgo \cite{schirgi_development_2024}) appeared only once each. Less than a fourth of the articles (n = 6, 22\%) used multiple LLM models to generate personas, for example, Midjourney and DALL-E to generate persona images \cite{salminen_picturing_2024}, GPT, Bard, and open source models to generate persona narratives \cite{schirgi_development_2024}, and using GPT, Claude, and Gemini to evaluate biases in personas \cite{gupta_evaluation_2024}. This indicates a limited degree of cross-model experimentation in current persona generation practices. Overall, the LLM usage in persona research is marked by evolution from early GPT-3 adoption in April 2023 to increased model diversification by late 2024 (see Figure \ref{fig:llms}), with GPT models maintaining dominance throughout the examined timeline.

\begin{figure}
    \centering
    \includegraphics[width=\linewidth]{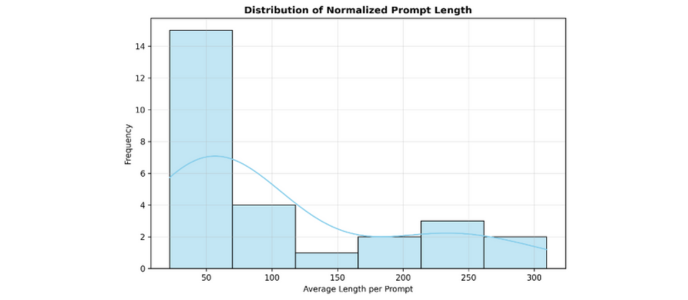}
    \caption{Prompt length (words) normalized by number of prompts in the entry.  We encourage the reader to zoom in if text is hard to read.}
    \label{fig:length distr}
\end{figure}

The complexity of persona prompts ranges from extremely simple one-liner prompts\footnote{For example, ``Pretend that you are a/an *persona*. Complete the following prompt in 3-5 sentences: When I think about *issue*, the first images or thoughts that come to mind are ...'' \cite{nguyen_simulating_2024}.} to extremely comprehensive and detailed, multi-stage prompting that underlies a complete persona generation system (e.g., \cite{choi_proxona_2024}), which is evident from the distribution of prompt word counts (see Figure \ref{fig:length distr}) and descriptive statistics of prompt length (M = 107.57, SD = 87.72, Mdn = 66, range: 287.50 words). The shortest prompt was 22 words and the longest 309 words.

On average, researchers used 3.1 persona prompts (SD = 2.9). The maximum number of prompts used was 12, with the minimum and mode being 1. The median number of persona prompts used was 2. More than half of the articles (n = 17, 62.96\%) used more than one prompt, thus opting for multi-prompt strategy. For example, Salminen et al. \cite{salminen_picturing_2024} used three prompts: (1) prompt to generate skeletal personas, (2) prompt to create an expanded full persona description, and (3) prompt to generate a persona picture. These prompts build on top of the previous step, in that the full persona expands the skeletal persona and the picture is based on the full persona description. Park et al. \cite{park_character-centric_2024} applied 11 persona prompts: 5 prompts for (a) persona generation (focused on different aspects, including image interpretation, trait specification and synthesis), 5 prompts for (b) guiding the interaction between persona and external conditions, and 1 prompt for (c) persona evaluation (contextual relevance scoring).

Impressively, Chen et al. \cite{chen_empathy-based_2024} reported 12 persona prompts, including (1) few-shot learning example for the generation of privacy attributes, (2) generating portrait image prompt, (3) generate device and browser, (4) generate persona description, (5) few-shot learning example for the generation of persona description, (6) generate privacy attributes, (7) few-shot learning example for the generation of schedule, (8) generate browsing history, (9) generate social media post content, (10) generate portrait image, (11) generate device and browser, and (12) generate schedule. Chen et al. \cite{chen_empathy-based_2024} illustrate well the versatility of using prompting at different stages of persona generation as well as the possibilities for downstream task application\footnote{For example, ``Provide ideas for this person to write posts (limit the word to 140 words) based on the profle [sic] and location history: \{profle\} \{schedule\}'' \cite{chen_empathy-based_2024}.}. Designing multiple persona prompts makes it possible to modularly combine these prompts to address different persona development tasks in automatic persona generation \cite{jung_personacraft_2025}. 

From the example of Chen et al. \cite{chen_empathy-based_2024}, one can also observe that certain prompts are designed to generate \textit{other} prompts (e.g., a prompt to generate the image generation prompt)---this can be referred to as embedded or circular design in prompt orchestration\footnote{Example from Chen et al. \cite{chen_empathy-based_2024} illustrating how outputs of different prompts are then used by other prompts in an embedded manner: ``Given the profle: \texttt{persona}, infer the browser and device the person uses: Prompt to generate schedule \texttt{<few-shot learning example>} You are acting as a game event designer. Write daily events for this person: \texttt{persona description}. Show me a reasonable schedule for this person from \texttt{start\_date} to \texttt{end\_date}. The life in the period is similar to 2021. You can generate fake but reasonable data that is related to the profle.''}. While the use of multiple prompts in the persona creation process leverages LLMs to a greater extent, increasing technical sophistication, it also makes the evaluation of the system by humans more complex, as individual prompts might cause cascading problems in downstream tasks or ``hidden pockets'' or issues that are challenging to detect. Instead of assessing the appropriateness (and outputs) of one prompt, humans are now required to verify multiple prompts and their outputs, while keeping the overall system and persona creation goals in mind. It is evident that this property of LLM usage makes persona evaluation more challenging than previously.

Less than a third (n = 8, 29.63\%) of the articles disclosed the hyperparameter values they used with the LLMs. 
This might be an indication of researchers merely applying the default hyperparameter values, without giving much thought to the influence of hyperparameter values on the output personas.


An interesting example of how LLMs support the integration of computational techniques in persona development is the high frequency of using dynamically inserted data values in the prompt, which took place in nearly three out of four (n = 20, 74.1\%) prompts, either using retrieval augmented generation (RAG) or some other technique\footnote{For example, ``Create personas of 5 each for $<$Role 1$>$, $<$Role 2$>$, and $<$Role 3$>$ at $<$Company$>$ in tables outlining names, genders, universities they attended, and brief descriptions about them.'' \cite{gupta_evaluation_2024}.}. In larger picture, integration data directly into the persona generation prompt paves way toward what Kim et al. \cite{kim_persona_2025} call ``computational personas''.

Finally, in five cases (18.52\%), researchers assigned a role for the LLM; ``role'' here does not refer to the LLM assuming the role of the persona but the role of a facilitator. These specified roles were (1) ``assistant to a social sciences researcher'' \cite{salminen_deus_2024,salminen_picturing_2024}, (2) ``you are a helpful and precise assistant for checking the quality of the AI assistant's responses in interactions'', (3) ``You are an assistant helping creators to improve their channels'' \cite{choi_proxona_2024}, and (4) ``role: system''.
In summary, researchers' use of LLMs ranged from individual prompts to system pipelines that carrying out multiple persona development tasks.

\subsection{RQ3: What Kind of Personas Do Researcher Generate with Persona Prompts?}


Researchers predominantly generate personas in text (n = 26, 96.3\%) and number (n = 18, 66.7\%) formats, while image generation for personas appears to have untapped potential due to its infrequent use (n = 2, 7.41\%) compared to these two other content formats. In other words, nearly all persona outputs involve text and two-thirds involve numbers. Numbers and text commonly appear together (n = 15, 55.56\%), while a fully ``rounded'' persona in terms of content that contains text, numbers, and image only appeared in two entries (7.41\%) (see Table \ref{tab:modality-combinations}). Because persona descriptions often involve text and images \cite{nielsen_template_2015}, it is not surprising that these content formats appear in persona prompts, although the rarity of persona image generation is surprising. The high prevalence of numbers is because, in most cases, personas were given an age \cite{schuller_generating_2024,salminen_deus_2024,de_winter_use_2024}. However, some researchers also use LLMs to \textit{predict} numerical persona attributes, such as sentiment score \cite{clements_innovative_2023}.


\begin{table}[htbp]
\centering
\caption{Content formats requested in persona prompts.}
\label{tab:modality-combinations}
\footnotesize
\begin{tabular}{p{2.5cm}cc}
\hline
\textbf{Format} & \textbf{Abs. Freq.} & \textbf{Rel. Freq.} \\
\hline
Text only & 9 & 33.33\% \\
Numbers only & 1 & 3.70\% \\
Text + numbers & 15 & 55.56\% \\
Text + numbers + image & 2 & 7.41\% \\
\hline
\end{tabular}
\end{table}

Most of the prompt entries (n = 17, 62.96\%) included instructions for the number of personas to generate. This took place either explicitly by specifying the number (...) or implicitly, most typically by referring to ``a persona'' (1 persona to generate) (e.g., \cite{schirgi_development_2024,paoli_improved_2023}). In most cases (n = 12, 70.6\%), the instruction was to create one persona. When not instructing to create one persona, the prompt entries asked the LLM to create a fairly high number of personas, including 15 \cite{gupta_evaluation_2024}, 20 \cite{de_winter_use_2024}, and 30 \cite{salminen_picturing_2024} personas. One entry asked the LLM to decide the appropriate number of personas to generate (``generate a minimum number of personas to represent the user data'' \cite{shin_understanding_2024}. Interestingly, a substantial number of prompt entries do not specify the number of personas to generate, even though this is an essential choice in persona development \cite{salminen_creating_2022}.

A substantial number (n = 11, 40.74\%) of the prompt entries involve instructions for the length of the persona output. The length requirements for personas varied from broad instructions\footnote{For example, ``one paragraph'' \cite{chen_why_2025}, ``brief descriptions'' \cite{gupta_evaluation_2024}, ``3-5 sentences'' \cite{nguyen_simulating_2024}.} to specific word or sentence requirements\footnote{For example, ``limit to 30 words'' \cite{chen_empathy-based_2024}, ``max 20 words'' (per persona description section) \cite{paoli_improved_2023}, ``not longer than 200 words'' \cite{benharrak_writer-defined_2024}.}. A major emphasis was put to curbing LLMs' verbosity\footnote{For example, ``Only respond with a single paragraph, you as a persona can only speak a maximum of 120 words'' \cite{choi_proxona_2024}, ``no longer than 4 lines ... max 200 words'' \cite{de_paoli_writing_2023}.}, which is somewhat in conflict with the goal of creating rich and detailed persona descriptions \cite{nielsen_template_2015}. On the other hand, length restrictions for the personas were often aligned with the goal of producing normalized output for analysis purposes\footnote{For example, ``Generate two separate sentences. Express in a simple way. Each sentence needs to be similar in length.''.}.

\begin{figure}
    \centering
    \includegraphics[width=0.8\linewidth]{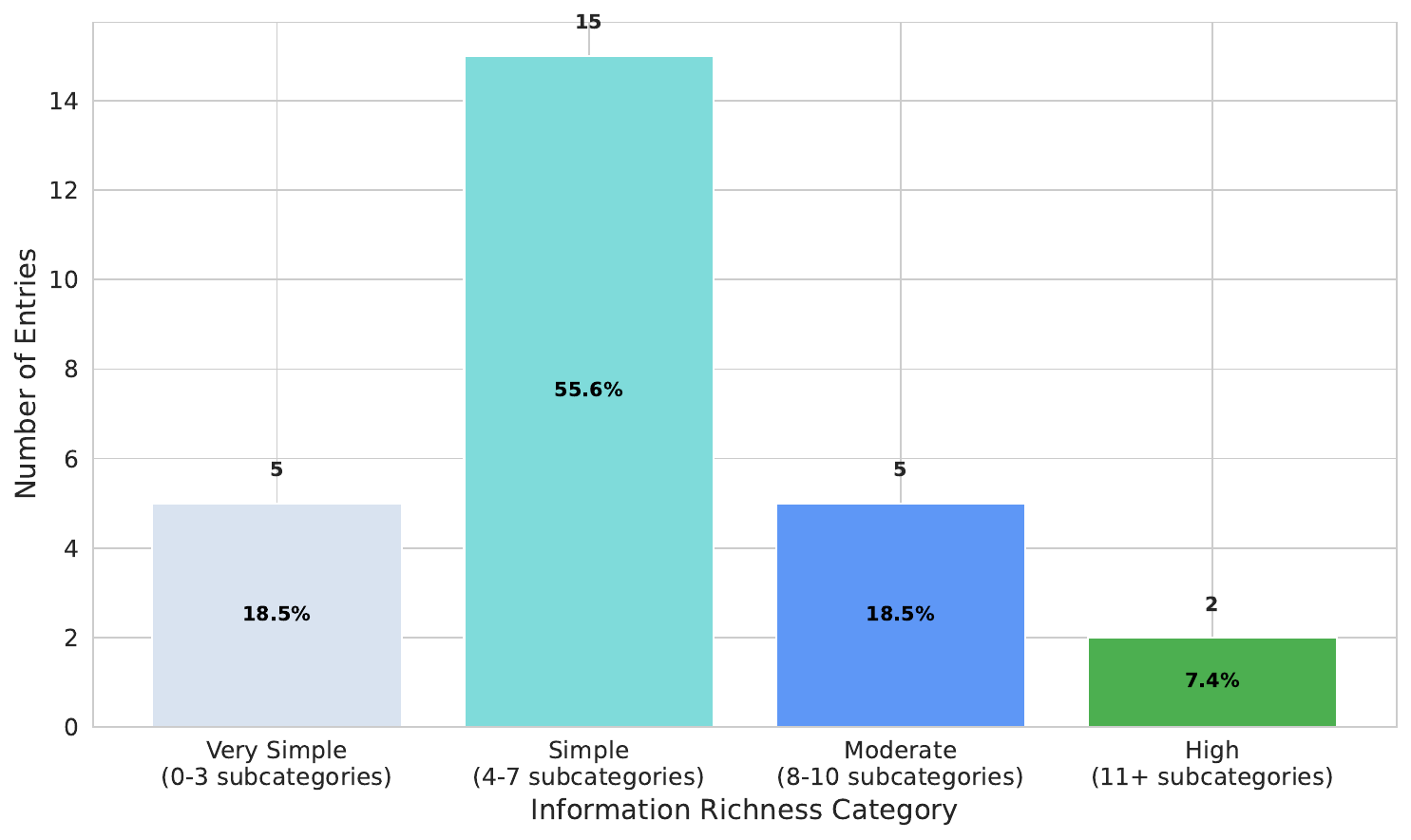}
    \caption{Persona information complexity when using LLM prompting to generate personas. Calculated based on the attributes requested in a persona prompt. Reference value ranges from Salminen et al. \cite{salminen_template_2020}. We encourage the reader to zoom in if text is hard to read.}
    \label{fig:information complexity}
\end{figure}

Persona attributes refer to the information present in persona descriptions \cite{nielsen_template_2015}. We listed, for each prompt entry, each explicit persona information attribute mentioned in the entry. On average, the entries included 5.48 information attributes (SD = 3.51), indicating that researchers generally use LLMs to generate relatively information-rich personas. However, previous research systematically analyzing data-driven persona templates found on average 8.83 information attributes in persona templates (SD = 2.57) \cite{salminen_template_2020}. Compared to these findings, our sample exhibited 38\% fewer information attributes per persona on average and 37\% higher variation in terms of SD. Using the reference values for persona information richness categorization provided by Salminen et al. \cite{salminen_template_2020}, the typical entry in our sample is situated in the ``Simple'' persona information richness category (4-7 persona attributes) (see Figure \ref{fig:information complexity}). So, LLM-generated personas appear to be slightly less informative than the previous generation of quantitative personas.

We also investigated what type of information the generated personas contain. For this, we developed five categories based on common persona information types (see, e.g., \cite{nielsen_template_2015}): (1) summary information that contains an overview of the persona, (2) demographics that contain basic demographic information, (3) behaviors that reflect the persona's behaviors, (4) attitudes that reflect the persona's attitudes, and (5) contextual information that is related to a specific context, such as research domain or industry (e.g., privacy).

The lead researcher manually extracted this information from each prompt entry. In most cases, the labels for the information were directly adopted from the prompts (e.g., if the prompt indicates ``age'' is a component of the desired persona output, then we adopt the label, ``age'' to describe this information); but in some cases, minor tweaks were made to merge the information into larger units (e.g., if one prompt indicated ``first name'' and ``last name'' separately and another indicated only ``name'', we adopted ``name'' as information label for both cases). The extraction yielded 150 instances of persona information mentioned in the prompts (see Table \ref{tab:categories}).

\begin{table}[htbp]
  \centering
  \begin{small}
  \caption{Category and Subcategories}
  \label{tab:categories}
  \begin{tabularx}{\columnwidth}{|p{2cm}|>{\raggedright\arraybackslash}X|}
    \hline
    \textbf{Category} & \textbf{Subcategories} \\
    \hline
    Summary (n = 18, 12.0\%) & background (8), description (4), details (2), index (2), current context (1), content (1) \\
    \hline
    Demographics (n = 78, 52.0\%) & age (16), name (12), occupation (11), gender (11), marital status (4), education (3), ethnicity (2), race (1), parental status (1), birthday (1), citizenship (1), religion (1), income (5), location (1), street (1), zip code (1), state (1), city (1), country (1), family environment (1), spoken language (1), living situation (1) \\
    \hline
    Behaviors (n = 18, 12.0\%) & interests (5), behavior (4), hobbies (3), typical day (1), digital literacy (1), IT skills (1), technology competence (1), religious activeness (1), friend count (1) \\
    \hline
    Attitudes (n = 19, 12.7\%) & personality (4), goals (2), needs (2), attitudes (2), expectations (1), sentiment (1), political ideology (1), political party (1), likes and dislikes (1), motivation for using civic services (1), plans for having children (1), tone of voice (1), style (1) \\
    \hline
    Contextual Information (n = 17, 11.3\%) & challenges (5), requirements (2), experiences (2), company (1), university (1), significant life events (1), dark secret (1), social status (1), appearance (1), role (1), potential engagement barriers (1) \\
    \hline
  \end{tabularx}
  \end{small}
\end{table}

\begin{figure}
    \centering
    \includegraphics[width=0.8\linewidth]{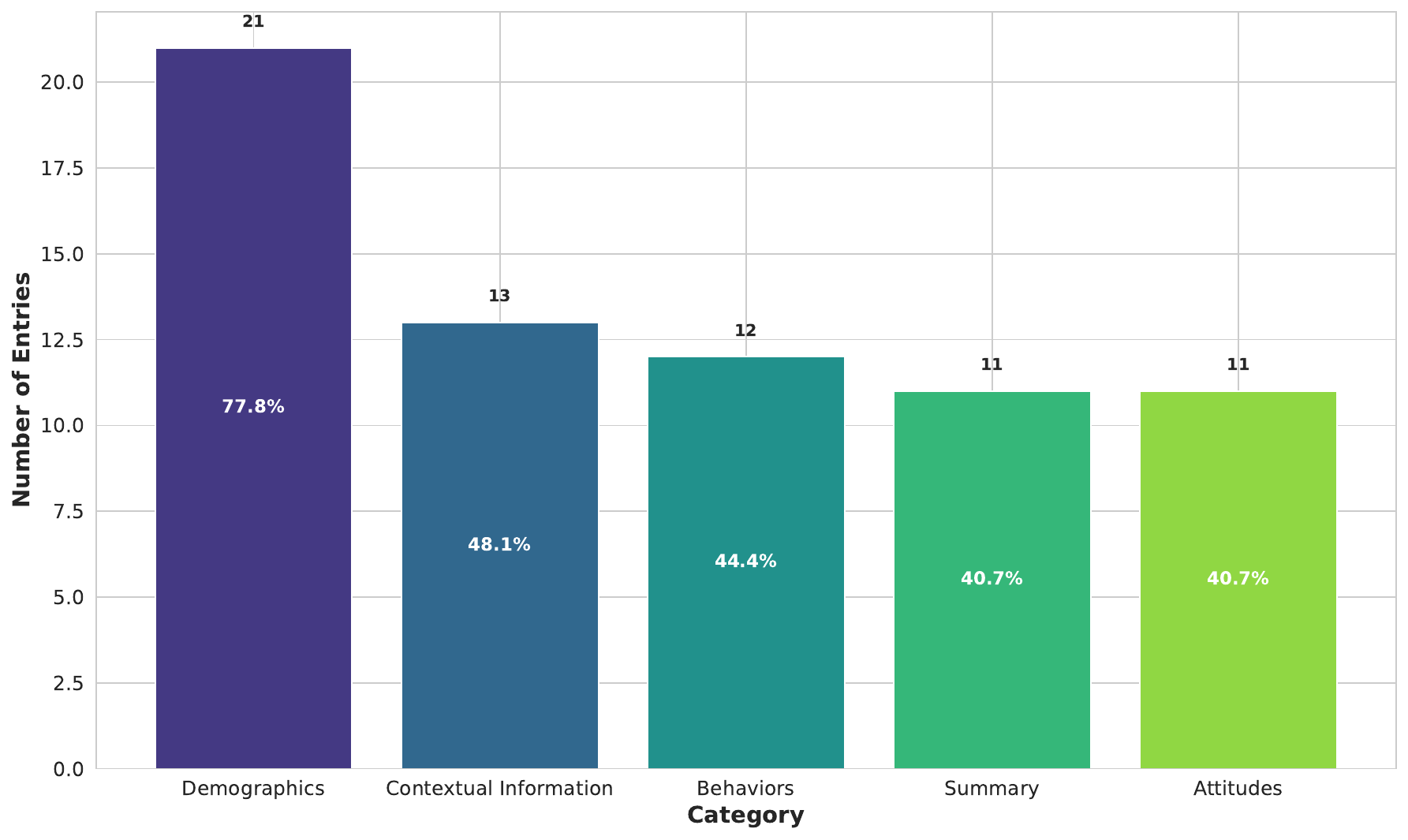}
    \caption{Frequency of persona prompt entries containing at least one subcategory from each main category for persona information. Note: An entry can include information multiple main categories.  We encourage the reader to zoom in if text is hard to read.}
    \label{fig:attributes}
\end{figure}

Results indicate that demographic information is the predominant category (n = 78, 52.0\%) of information that persona creators include in the prompt (see Table \ref{tab:categories}). We also computed how many prompt entries included at least one information attribute from each main information category. These results (see Figure \ref{fig:attributes}) indicate that the vast majority of persona prompt entries included demographics (n = 21, 77.78\%), making it the most frequently occurring category for persona information. Contextual Information was included in around half of the entries (n = 13, 48.15\%). The rest of the persona information categories occurred in similar frequencies (see Figure \ref{fig:attributes}).

The predominance of demographic information is consistent with previous reviews both traditional \cite{nielsen_template_2015} and data-driven \cite{salminen_template_2020} persona templates, in which demographics has been identified as central information. However, there is also research investigating the negative effects of demographics in terms of stereotyping; in any case, demographics seem to remain a key component in personas even in LLM-generated personas. In addition to the commonness of demographic information, other traditional user representation information, including user behaviors, attitudes, and contextual information remain reasonably prevalent also in LLM-generated personas, implying that practices for creating personas are carried over to the new technological environment.

\section{Discussion} \label{discussion}

\subsection{Implications for Persona Research}


The use of LLMs in persona generation is a logical continuum to research on data-driven personas \cite{salminen_survey_2021}, which applies novel computational techniques to support persona generation workflows \cite{shin_understanding_2024} and tasks involved in persona creation \cite{jung_personacraft_2025}. Based on our findings, it seems that some traditions from ``classic'' persona development \cite{cooper_inmates_1999,nielsen_template_2015} have carried over to LLM-generated personas, visible from the persona attributes requested, which contain traditional attributes, such as demographics, behaviors, and attitudes, often combined in textual and numerical formats (as done in classic persona descriptions that contain text and numbers). Additionally, consistent with the notion of mental modeling \cite{nielsen_personas_2019}, LLMs are frequently asked to provide predictions from the persona's perspective.

However, there also exist deviations from the classic persona creation traditions. First, researchers often aim to control the verbosity of the LLM in order to generate concise persona descriptions. This is different from the idea of creating rounded and detailed persona narratives \cite{nielsen_personas_2019}. Also, while images are an integral part of classic persona descriptions \cite{nielsen_template_2015}, LLMs are rarely prompted to generate persona pictures. Perhaps more importantly, we can observe two interesting trends, which are (a) the inclusion of data and/or variables in the persona generation prompt and (b) requesting the persona information in a structured data format. The former is interesting because, traditionally, data analysis and persona write-up have been separate parts of the persona development process \cite{nielsen_personas_2019}. LLMs now enable adding data in conjunction with providing instructions for persona generation; this process certainly entails both opportunities and risks for persona generation, such as decreasing transparency and human agency \cite{prpa_challenges_2024}. This lack of human involvement is also visible from the fact that researchers have started chaining and combining persona prompts; evaluating such systems is extremely challenging because different cause-and-effect connections between different prompts may not be immediately apparent and understanding complex systems requires a lot of effort and expertise. Therefore, considerable research is needed on evaluating both (a) individual persona prompts (and their outputs) and (b) whole systems consisting of multiple, interrelated persona prompts.

Similarly, the frequent use of structured JSON outputs (50\% of cases requiring a structured format) indicates researchers are treating personas as data objects rather than narrative tools for empathy building. Most prompts generate single personas instead of diverse persona sets, which limits their ability to represent the full spectrum of a user population. The dominance of GPT models (76\% of instances) without cross-model comparison raises questions about whether researchers are getting the best possible persona outputs or simply using the most familiar tool.

\subsection{Practical Recommendations}

There is the risk that persona prompts break the principle of `data-drive' in persona development, replacing real user data with LLMs' creative abilities. This applies specifically to cases in which personas are generated from the so-called ``knowledge'' of the LLM without using any primary data about the users. To keep LLM-generated personas data-driven, we recommend including primary data about users in persona prompts, either in aggregated or individual level. 

Another practical recommendation is to familiarize oneself with persona theory. Some persona prompts appear ``shallow'' in the sense that they do not exhibit detailed understanding of persona theory, such as considering empathy creation, perspective taking, and representativeness and diversity. There may be a certain drift in which computationally oriented researchers lack some information from the qualitative persona creation tradition; reading more about persona theory--e.g., \cite{cooper_inmates_1999,nielsen_personas_2019} can be helpful in this regard, as well as learning about the principles of data-driven personas \cite{jansen_data-driven_2021}.

\subsection{Limitations and Future Work}

The study has some limitations. First, the field is evolving rapidly. The findings we present here may stale over time, requiring updated analyses and revisits. Second, our analysis focuses on elementary variables; future research could delve deeper into more theoretically informed analyses, such as categorizing the persona prompts' technical complexity, and assessing if and how they consider algorithmic fairness guidelines (a.k.a. guardrails). Third, because we do not present evidence on empirical relationships between prompt design and outputs, we are unable to articulate what  the best practices for persona design would be from an evidence-based perspective; future work is needed. 
Because we make the persona prompt collection publicly available, other research has a good opportunity to build on top of and expand on our findings.

\bibliography{references}

\end{document}